\documentclass[prl,reprint,superscriptaddress,showpacs,showkeys,longbibliography]{revtex4-1}

\pdfoutput=1
\usepackage{siunitx}

\usepackage{amsmath,amsfonts,amssymb,bm,graphicx,color,xcolor,fancyhdr,multirow}
\usepackage[percent]{overpic}
\usepackage{subscript}

\usepackage[colorlinks=true,linkcolor=blue,urlcolor=blue,citecolor=blue]{hyperref}

\newcommand{\Molek}[2]{{#1\textsubscript{{#2}}}}

\newcommand{\bLAB}{$\beta$-\Molek{Lu}{}\Molek{Al}{}\Molek{B}{4}}

\newcommand{\aYAB}{$\alpha$-\Molek{Yb}{}\Molek{Al}{}\Molek{B}{4}}
\newcommand{\bYAB}{$\beta$-\Molek{Yb}{}\Molek{Al}{}\Molek{B}{4}}

\begin{document}
 
\title{Strong in-plane anisotropy in the electronic structure of fixed-valence \bLAB}

\author{Pascal Reiss}	% Checked
\email[]{pascal.reiss@physics.ox.ac.uk}
\affiliation{Cavendish Laboratory, University of Cambridge, Cambridge, CB3 0HE, United Kingdom}
\affiliation{Clarendon Laboratory, University of Oxford, Oxford, OX1 3PU, United Kingdom}

\author{Jordan Baglo}	% Checked
\altaffiliation[Current address: ]{Institut Quantique, D\'epartement de Physique \& RQMP, Universit\'e de Sherbrooke, Sherbrooke, Qu\'ebec J1K 2R1, Canada}
\affiliation{Cavendish Laboratory, University of Cambridge, Cambridge, CB3 0HE, United Kingdom}

\author{Hong'En Tan}	% Not found
\affiliation{Cavendish Laboratory, University of Cambridge, Cambridge, CB3 0HE, United Kingdom}
\affiliation{Complex Systems Group, Institute of High Performance Computing, A*STAR, Singapore, 138632}

\author{Xiaoye Chen}	% Not found
\affiliation{Cavendish Laboratory, University of Cambridge, Cambridge, CB3 0HE, United Kingdom}

% \author{Hisatomo Harima}	% Checked
% \affiliation{Kobe University, Kobe 657-8501, Japan}

\author{Sven Friedemann}	% Checked
\affiliation{HH Wills Laboratory, University of Bristol, Bristol, BS8 1TL, United Kingdom}

% \author{Swee K. Goh}	% Checked
% \affiliation{Cavendish Laboratory, University of Cambridge, Cambridge, CB3 0HE, United Kingdom}
% \affiliation{Chinese University of Hong Kong, Shatin, N.T., Hong Kong}

\author{Kentaro Kuga}	% Checked
\altaffiliation[Current address: ]{Toyota Technological Institute, Nagoya, 468-8511, Japan}
\affiliation{Institute for Solid State Physics, University of Tokyo, Kashiwa 277-8581, Japan}
% \affiliation{RIKEN SPring-8 Center, Sayo-gun, Hyogo 679-5148, Japan}

\author{F. Malte Grosche}	% Checked
% \email[]{fmg12@cam.ac.uk}
\affiliation{Cavendish Laboratory, University of Cambridge, Cambridge, CB3 0HE, United Kingdom}

\author{Satoru Nakatsuji}	% Checked
\affiliation{Institute for Solid State Physics, University of Tokyo, Kashiwa 277-8581, Japan}
\affiliation{CREST, Japan Science and Technology Agency, Kawaguchi, Saitama 332-0012, Japan}
\affiliation{Department of Physics, University of Tokyo, Bunkyo-ku, Tokyo 113-0033, Japan}
\affiliation{Trans-scale Quantum Science Institute, University of Tokyo, Bunkyo-ku, Tokyo 113-0033, Japan}

\author{Michael Sutherland}	% Checked
\email[]{mls41@cam.ac.uk}
\affiliation{Cavendish Laboratory, University of Cambridge, Cambridge, CB3 0HE, United Kingdom}

\date{\today}

\begin{abstract}
% The origin of intrinsic quantum criticality in the heavy-fermion superconductor \bYAB\ has been a puzzling question since its first observation. Strong Yb valence fluctuations and/or the peculiar crystal structure may play a dominant role in producing a nodal band structure and a topological phase transition of the Fermi surface. Here we present quantum oscillation measurements and electronic structure calculations on the isostructural, but non-critical reference compound \bLAB\ in the absence of valence fluctuations. The very convincing agreement between theory and experiment reveals a Fermi surface of \bLAB, which differs greatly from that of \bYAB, and which suggests that the Yb 4$f$ states contribute to the Fermi surface volume there. Moreover, we find that electronic structure shows a marked anisotropy within the ($ab$) plane, due to the crystal structure and deviations from ideal heptagonal boron rings. Our findings suggest that a full account of quantum criticality in \bYAB\ should include this intrinsic anisotropy. 
The origin of intrinsic quantum criticality in the heavy-fermion superconductor \bYAB\ has been attributed to strong Yb valence fluctuations and to its peculiar crystal structure. Here, we assess these contributions individually by studying the isostructural but fixed-valence compound \bLAB. Quantum oscillation measurements and DFT calculations reveal a Fermi surface markedly different from that of \bYAB, consistent with a `large‘ Fermi surface there. We also find an unexpected in-plane anisotropy of the electronic structure, in contrast to the isotropic Kondo hybrization in \bYAB.
\end{abstract}


% \keywords{}

\maketitle

% \paragraph*{Introduction.}

At a quantum critical point (QCP), a continuous, zero-temperature phase transition occurs when the ordered and disordered phases are energetically degenerate. Usually, this requires tuning materials to some material-de\-pen\-dent critical pressure, composition or magnetic field \cite{Lohneysen2007}. However, there are rare case of `intrinsic' QCPs which occur without tuning. Studying intrinsic QCPs aims at identifying robust mechanisms for quantum criticality, rising the prospect to link them to strange electronic phases and unconventional superconductivity \cite{Esterlis2019,Hauck2020}.

The heavy-fermion superconductor \bYAB\ is an example for such an intrinsically quantum critical system \cite{Nakatsuji2008}. Initial evidence for its non-accidental nature was the observation of quantum critical behavior in this stoichiometric, clean system at zero magnetic field, which was later extended to finite pressures, providing an example for a quantum critical phase \cite{Matsumoto2011,Tomita2015a}. Transport and thermodynamic measurements identified non-Fermi liquid behavior, yet the Wiedemann-Franz law is obeyed, demonstrating that quasiparticles remain intact \cite{Nakatsuji2008,Nakatsuji2010,Sutherland2014}. 

\begin{figure*}
	\begin{overpic}[width=18cm]{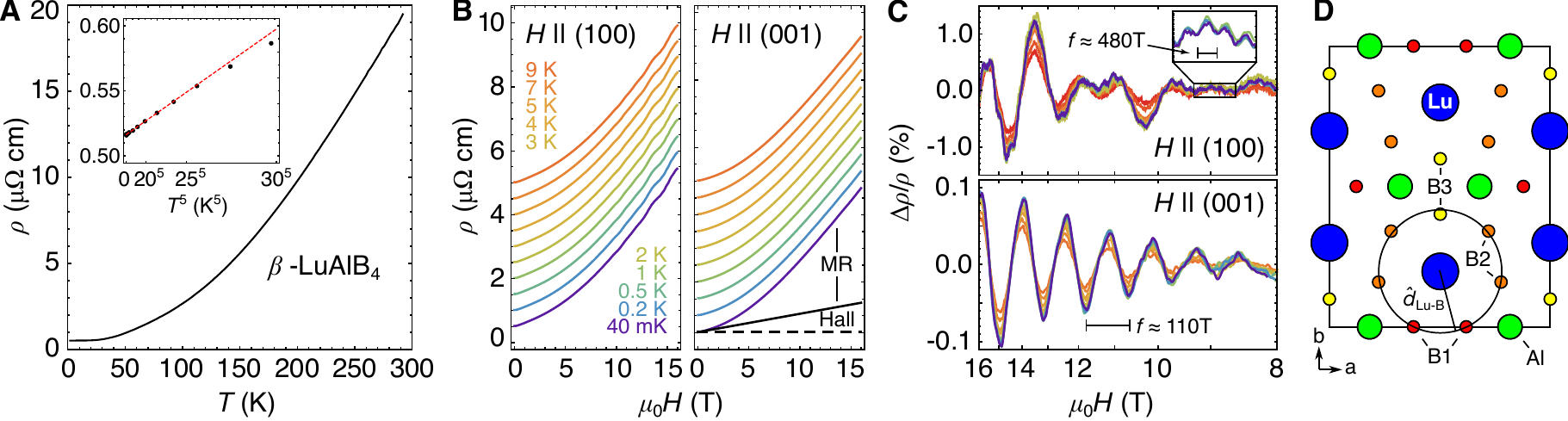}
	\end{overpic}
	\caption{
		Temperature (A) and magnetic field dependence (B) of the electrical resistivity of single-crystalline \bLAB. The insert shows the low-temperature resistivity against $T^5$, and the red dashed line is a linear fit below \SI{25}{\kelvin}. Data in (B) are vertically offset for clarity. 
		(C) Background subtracted data plotted against inverse field $1/H$. The periodic oscillations in $\Delta \rho / \rho(T)$ are clearly seen.
		(D) A 2D projection of the crystal structure. The seven-fold B ring is highlighted, consisting of atomic sites B1-B3. The average Lu-B distance is given as $\hat d_\text{Lu-B}$ which shows small variations, most pronounced for B3 \cite{Macaluso2007}.}
	\label{fig:coolDown}
\end{figure*}

It was realised that the peculiar crystal environment of the Yb atoms, as well as strong Yb valence fluctuations are central for understanding the physics of \bYAB\ \cite{Okawa2010,Watanabe2010,Nevidomskyy2008}. The material crystallizes into the \Molek{Th}{}\Molek{Mo}{}\Molek{B}{4} structure where a distorted honeycomb lattice of Yb and Al atoms is sandwiched between layers of B atoms. As shown in Fig.~\ref{fig:coolDown}, these B atoms occupy three independent sites B1--B3, which give rise to a network of five- and sevenfold boron rings \cite{Macaluso2007}. The Kondo hybridization induced by the orbital overlap within the Yb-B-ring structure was argued to be strongly anisotropic between the crystal $c$ axis and the ($ab$) plane, but \emph{isotropic} within the ($ab$) plane \cite{Ramires2012}. This hybridization can induce the observed $T$/$B$ scaling of the magnetic susceptibility, and can also explain the intrinsic quantum criticality in \bYAB\ \cite{Ramires2012,Matsumoto2011,Holanda2011,Watanabe2016,Watanabe2019,Ramires2014,Komijani2018}. Recently, strong evidence supporting this in-plane isotropy was found in the linear dichroism of core-level x-ray photoemission and NMR studies \cite{Kuga2019,Takano2016}.

The isomorph \aYAB\ features the same local symmetry and similar valence fluctuations of the Yb atoms, as well as an essentially identical in-plane isotropy of the Kondo hybridization \cite{Macaluso2007,Okawa2010,Kuga2018,Kuga2019}. However, it does not show superconductivity nor intrinsic quantum criticality %, and its low-temperature Sommerfeld coefficient is only about half that of \bYAB
\cite{Macaluso2007}. This indicates that the effect of the lattice cannot be truncated to the local environment of the Yb atoms, and/or that subtle differences in the Yb-B structures between $\alpha$- and \bYAB\ are pivotal \cite{Takano2016,Kuga2018}. 

It is therefore paramount to assess the importance of the lattice structure in detail \cite{Friedemann2013,Reiss2013}. For this purpose, \bLAB\ is ideally suited, as it crystallizes into the same structure as \bYAB, with nearly identical lattice parameters \cite{Macaluso2007,Seitz2007}. The Lu 4$f$ states are fully occupied and fall well below the Fermi level. Thus, the electronic structure of \bLAB\ represents the `small' Fermi surface limit of \bYAB, and the Kondo effect and valence fluctuations are suppressed. 

In this Letter, we investigate the electronic structure of \bLAB\ using Shubnikov-de Haas oscillation measurements. We resolve all Fermi surface sheets predicted by DFT calculations and we extract the charge carrier masses. Our findings are broadly consistent with DFT predictions, validating a closer look at the electronic structure in real space. This reveals that the Lu-B hybridization and the charge distribution in the unit cell are much more \emph{anisotropic} than previously thought, in contrast to the \emph{isotropic} Kondo hybridization in \bYAB.

Using the \Molek{Al}{} flux method \cite{Macaluso2007}, thin ($d \lesssim \SI{10}{\micro\meter}$), plate-like single crystals of \bLAB\ were grown, with the crystal $c$ axis normal to the plates, confirmed by X-ray diffraction measurements. For the resistivity measurements, we used a standard four contact setup with $I < \SI{1}{\milli\ampere} || (ab)$. Contacts were spot-welded and fixed with DuPont 6838 silver epoxy. The irregular shape of the brittle samples as well as the large relative uncertainty in the sample thickness put constraints on the contact layout and the determination of absolute values of the resistivity (cf.~the Supplemental Material \footnote{See Supplemental Material at [URL will be inserted by publisher] for further field measurements and analyses, and a summary table}). % Accounting for a large (relative) uncertainty of the sample thickness, our best estimate for the residual resistivity at lowest temperatures is $\rho_0 \approx \SI{0.52}{\micro\ohm\centi\meter}$. % as shown in Figure~\ref{fig:coolDown}A. % Below \SI{4}{\kelvin}, a minute increase in $\rho$ could be observed with $\Delta \rho_0 / \rho_0 < 0.03\%$ (not shown) which we attribute to small impurity scattering. 

Several samples were screened using a Quantum Design PPMS, and the best sample with a residual resistivity ratio RRR$ = 34$ was used for all experiments. Quantum oscillation experiments were conducted using an \SI{18}{\tesla} dilution refrigerator in Cambridge equipped with low-temperature transformers. Two sets of measurements were performed, spanning sample rotations by almost \SI{90}{\degree} with the magnetic field ranging from the crystal (001)-axis to the (100)-axis and to the (010)-axis.

Band structure calculations were carried out using density functional theory, linear augmented plane waves and the PDE generalised gradient approximation as implemented in WIEN2k \cite{Perdew1996,Schwarz2003}. Relativistic local orbits and spin-orbit coupling were included. Small Fermi surface pockets were resolved using 100,000 $k$-points in the Brillouin zone. $RK_{\textbf{max}}$ was set to $7.0$, and experimental lattice parameters and atomic positions were used \cite{Macaluso2007}. Fermi surfaces were plotted using XCrySDen, and extremal orbits extracted using SKEAF \cite{xcrysden,Rourke2012}.

% \paragraph*{Results.}

In Figure~\ref{fig:coolDown}A, we show the sample resistivity $\rho(T)$ as a function of temperature $T$. Good metallic behavior is found, and no phase transitions including superconductivity could be identified above $T \approx \SI{40}{\milli\kelvin}$, in agreement with previous reports \cite{Macaluso2007}. The small residual resistivity $\rho_0 \approx \SI{0.52}{\micro\ohm\centi\meter}$ in the zero-temperature limit suggests large carrier densities and/or mobilities. As shown in the inset of Fig.~\ref{fig:coolDown}A, $\rho(T)$ shows a marked $T^5$ dependence below $T \approx \SI{25}{\kelvin}$. This demonstrates that electron-electron interactions are not dominating the quasiparticle scattering in this regime, in contrast to (non-critical) strongly-correlated electron systems, where electron-electron scattering leads to a Fermi liquid $T^2$ behavior. The $T^5$ dependence observed here is an indicator for dominant quasi-elastic electron-phonon scattering instead.

% \begin{figure}
% 	\begin{overpic}[width=8cm]{Fig1A.pdf}
% 	\end{overpic}
% 	\caption{Magnetoresistance and quantum oscillation measurements in \bLAB\ as a function of temperature and field orientation. (A) Raw magnetoresistance for two selected field orientations as a function of temperature. Note the visible quantum oscillations for $H || (100)$ in the raw data at high fields. Data are vertically offset for clarity. (B) Polynomial background subtracted data plotted against inverse field $1/H$. The periodic oscillations in $\Delta \rho / \rho(T)$ are clearly seen for all orientations.}
% 	\label{fig:dataInField}
% \end{figure}

\begin{figure*}
	\begin{overpic}[width=18cm]{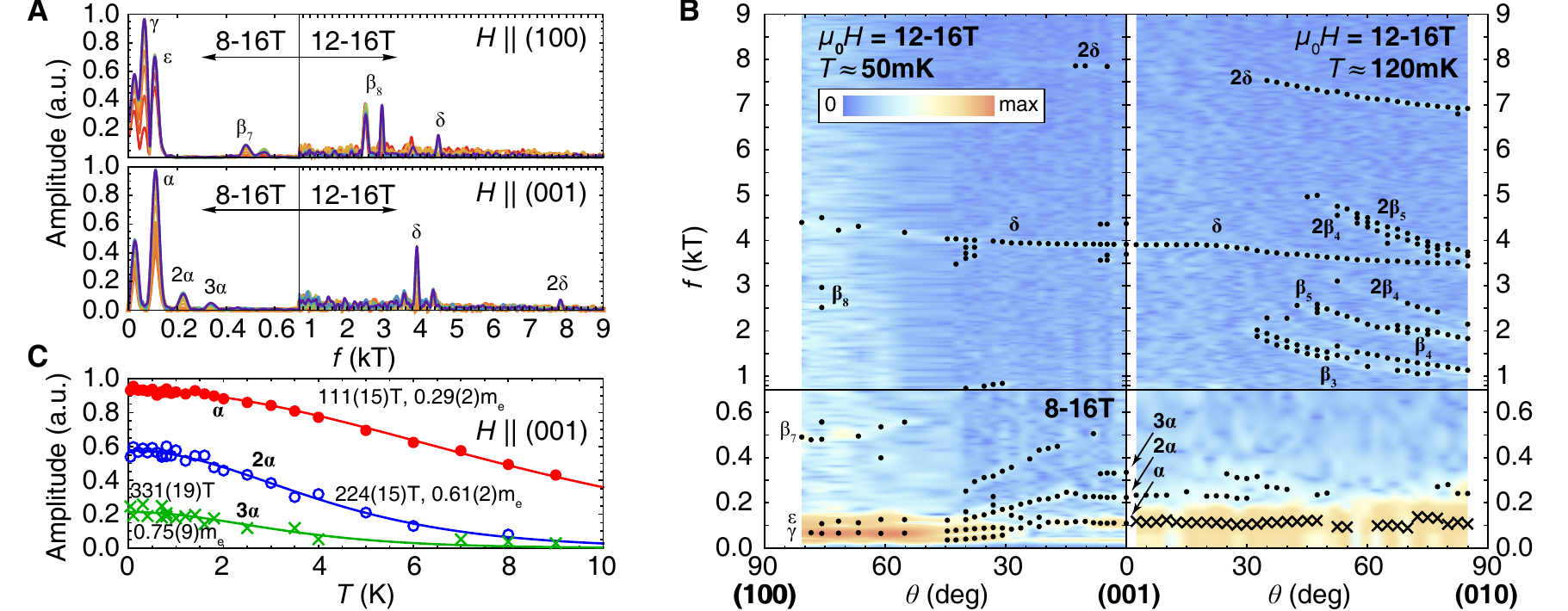}
	\end{overpic}
	\caption{Quantum oscillations in \bLAB. (A) FFT spectra for two field orientations, showing a range of peaks. (B) Angular dependence of the FFT spectral intensity shown as a color map. Extracted peak positions are shown as black dots, and frequencies extracted from direct fits are shown as black crosses (cf. Ref.~\citenum{Note1}). (C) Temperature dependence of the amplitude of the $\alpha$, $2\alpha$ and $3\alpha$ oscillations (points), and a fit to the temperature dependence of the Lifshitz-Kosevich form (solid lines).}
	\label{fig:qoAnalysis}
\end{figure*}

To determine the electronic structure, we performed electrical transport measurements up to $\mu_0 H = \SI{18}{\tesla}$ for various temperatures and field orientations. Figure~\ref{fig:coolDown}B shows a selection of resistivity measurements, from which several aspects can readily be inferred: Firstly, we observe a finite slope $\text{d}\rho / \text{d}H$ in the zero-field limit, which we attribute to the Hall effect visible due to some voltage contacts misalignment on the sample. Our DFT calculations, discussed below and in agreement with previous reports \cite{OFarrell2009}, predict that \bLAB\ is a compensated two-band system. If both hole- and electron-like charge-carriers had the same mobility, no Hall component would be expected ($R_H = 0$). The finite $R_H$ observed here shows that the mobilities differ for both carriers.

Secondly, we note that the magnetoresistance is very large. Separating the Hall signal from the longitudinal resistivity as demonstrated in Fig.~\ref{fig:coolDown}B, we find that the latter increases almost eight-fold in $\SI{16}{\tesla}$. From a two-band model fit shown in Fig.~SM1 \cite{Note1}, we extract the quasiparticle mobilities  $|\mu_1| \approx \SI{2600}{\centi\meter^2 / \volt\second}$ and $|\mu_2| \approx \SI{1600}{\centi\meter^2 / \volt\second}$, which are similar to those attributed to the high-mobility carriers detected in \bYAB\ below the coherence temperature \cite{OFarrell2012}. They also broadly match the values extracted from quantum oscillations and are consistent with predictions of our DFT calculations \cite{Note1}.

Thirdly, in particular for $H || (100)$, strong quantum oscillations can readily be seen in Fig.~\ref{fig:coolDown}B, demonstrating the high quality of the sample. The quantum oscillations remain visible even for $T = \SI{9}{\kelvin}$, suggesting small quasiparticle masses $m$, which is consistent with the already inferred large quasiparticle mobilities $|\mu| \sim 1/m$.

We now turn to a quantitative analysis of these quantum oscillations. In Fig.~\ref{fig:coolDown}C, a smooth, low-order polynomial background has been removed. When plotted against $1/H$, multiple frequencies $f$ are apparent, which correspond to extremal cross-sectional areas $A$ of the Fermi surface perpendicular to the magnetic field \cite{Shoenberg1984} $f = A \hbar / (2 \pi e)$. We employ an FFT, shown in Fig.~\ref{fig:qoAnalysis}A, to extract the oscillation frequencies and amplitudes as a function of temperature and field orientation (see also Ref.~\citenum{Note1}). All obtained frequencies are shown in Fig.~\ref{fig:qoAnalysis}B on top of the FFT spectra, represented as a background color scale. We identify a set of branches for fields close to the (010) direction, which we refer to as $\beta$. These branches display an approximate $1/\sin(\theta)$ dependence, where $\theta$ measures the angle away from (001). Such a dependence is consistent with a weakly warped two-dimensional Fermi surface pocket along the reciprocal $b$ axis. Next, we find two weakly orientation-dependent lines for $\SI{3.4}{\kilo\tesla} \le f \le \SI{4.5}{\kilo\tesla}$ and $f \ge \SI{6.8}{\kilo\tesla}$ which we call $\delta$ and $2\delta$. Their flat dependencies are consistent with nearly spherical Fermi surface pockets, however the integer ratio between their frequencies suggests that $2\delta$ is the first harmonic of $\delta$. Finally, multiple frequencies below $\SI{0.9}{\kilo\tesla}$ are found with no obvious analytical dependence.

\begin{figure*}
	\begin{overpic}[width=180mm]{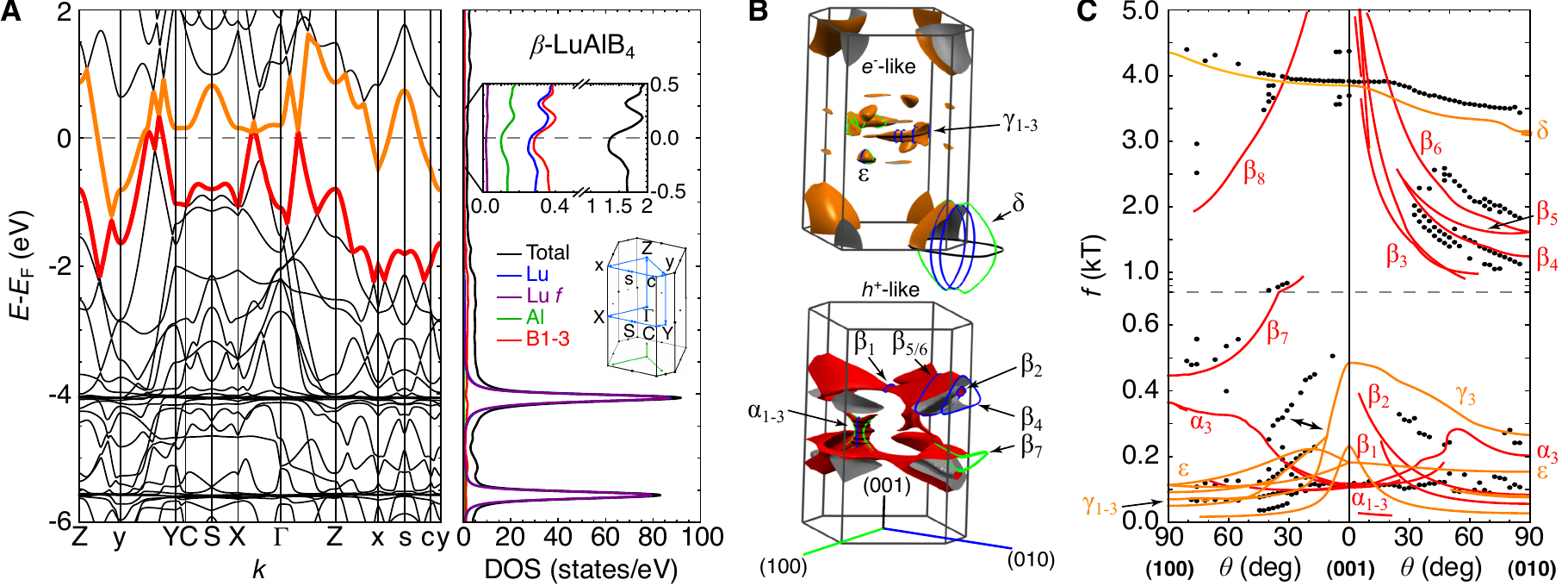}
	\end{overpic}
	\caption{DFT simulations and comparison to experiment. (A) Band structure and the density of states (DOS). Insets show the magnified DOS at the Fermi level, and the path in $k$ space. (B) Fermi surface sheets corresponding to the bands crossing the Fermi level. Orbits surrounding extremal cross sections are shown for field directions along the reciprocal axes (color coding as shown). (C) Comparison of predicted frequencies (lines) with extracted fundamental ones (dots).}
	\label{fig:DFTResult}
\end{figure*}

Next, cyclotron masses are extracted by following the oscillation amplitudes as a function of temperature, as shown in Fig.~\ref{fig:qoAnalysis}C. Fitting the Lifshitz-Kosevich form \cite{Shoenberg1984}, we find $m \approx 0.3m_e$ for $\alpha$, which is much smaller than any cyclotron mass reported on the more strongly correlated \bYAB, as expected \cite{OFarrell2009}. The integer ratios between the frequencies and masses of orbits $\alpha$, $2\alpha$ and $3\alpha$ allow us to identify $\alpha$ as a fundamental frequency, and $2\alpha$ and $3\alpha$ as harmonics. Further orbits are discussed in Ref.~\citenum{Note1}. From now, we will omit harmonics identified in this way, as summarised in Fig.~\ref{fig:DFTResult}C.

To better understand the observed quantum oscillations, we now turn to our DFT calculations. In Fig.~\ref{fig:DFTResult}A, we show the computed band structure and density of states (DOS) over a wide energy range. The localized Lu 4$f$ states are easily identified about \SI{4.1}{e\volt} and \SI{5.6}{e\volt} below the Fermi level $E_\text{F}$. In the vicinity of $E_\text{F}$, the DOS is dominated by Lu and B states whereas the Al states play a minor role. Two bands cross the Fermi level, which give rise to one hole-like and one electron-like Fermi surface sheet, as shown in Fig.~\ref{fig:DFTResult}B. The hole-like sheet consists of two strongly warped cylinders along (010) which give rise to the extremal orbits $\beta_{1-8}$. They are joined by two three-dimensional pockets, producing orbits $\alpha_{1-3}$. The electron-like sheet consists of nearly spherical pockets in the corners of the first Brillouin zone which are the origin of the orbit $\delta$. Further pockets with varying sizes surround the $\Gamma$ point, with orbits labelled $\gamma_{1-3}$ and $\epsilon$.

In Fig.~\ref{fig:DFTResult}C, we compare the computed extremal Fermi surface cross-sections translated into quantum oscillation frequencies to the experimentally determined fundamental frequencies. The agreement is convincing: the line $\delta$ and the branches $\beta_{3-6}$ are reproduced, and most frequencies below $\SI{0.9}{\kilo\tesla}$ can be mapped within experimental resolution ($\Delta f \approx \SI{20}{}-\SI{50}{\tesla}$ depending on field range). DFT slightly underestimates the size of the large Fermi surface pockets, i.e. predicted frequencies are slightly smaller than observed ones, as evident from Fig.~\ref{fig:DFTResult}C. This applies to both electron-like and hole-like bands, which leaves the overall charge balance unchanged.

%\begin{figure}
%	\begin{overpic}[width=80mm]{Fig4.pdf}
%	\end{overpic}
%	\caption{Inequivalence of the Boron atoms. (A) Charge carrier distribution $\rho_\text{el}$ in the proximity of the Lu atom and the sevenfold B ring. Electronic states within $\Delta E = k_\text{B} \cdot \SI{300}{\kelvin}$ of the Fermi level were included. The closed surfaces encapsulate regions with $\rho_\text{el} > 5.5 \cdot 10^{26}e/m^3$. (B) the charge carrier distribution in the Lu/Al layer of the unit cell, with the B atoms projected. (C) Comparison of the density of states contributions of the B atoms. Upper panel: total contribution, lower panel: $p_z$ states only. }
%	\label{fig:BoronAnisotropy}
%\end{figure}

% \paragraph*{Discussion.}
Having determined the electronic structure of \bLAB, the isostructural but fixed-valence reference compound to \bYAB, we now discuss the two main implications of the findings. Firstly, the frequencies of the quantum oscillations and their angular dependence, as well as the cyclotron masses reported here differ greatly from results on \bYAB\ studied previously \cite{OFarrell2009}. The most notable difference is the absence of the spherical pocket $\delta$ in \bYAB. Conversely, large cross-sectional areas were observed in \bYAB\ (dubbed $\beta$ in Ref.~\citenum{OFarrell2009}), which have no counterpart in \bLAB. Such substantial differences in the Fermi surface suggest that in the presence of large magnetic fields, the Fermi surface of \bYAB\ is not `small' but `large' and the 4$f$ hole is delocalized \cite{OFarrell2009}. Moreover, the experimental cyclotron masses of \bLAB\ are in good agreement with predicted values (Table~SMI), but they are on average a factor 10 smaller than in \bYAB\ \cite{OFarrell2009}. This reflects the significant mass enhancement due to the presence of correlated $f$-states near the Fermi energy.

\begin{figure*}
	\begin{overpic}[width=180mm]{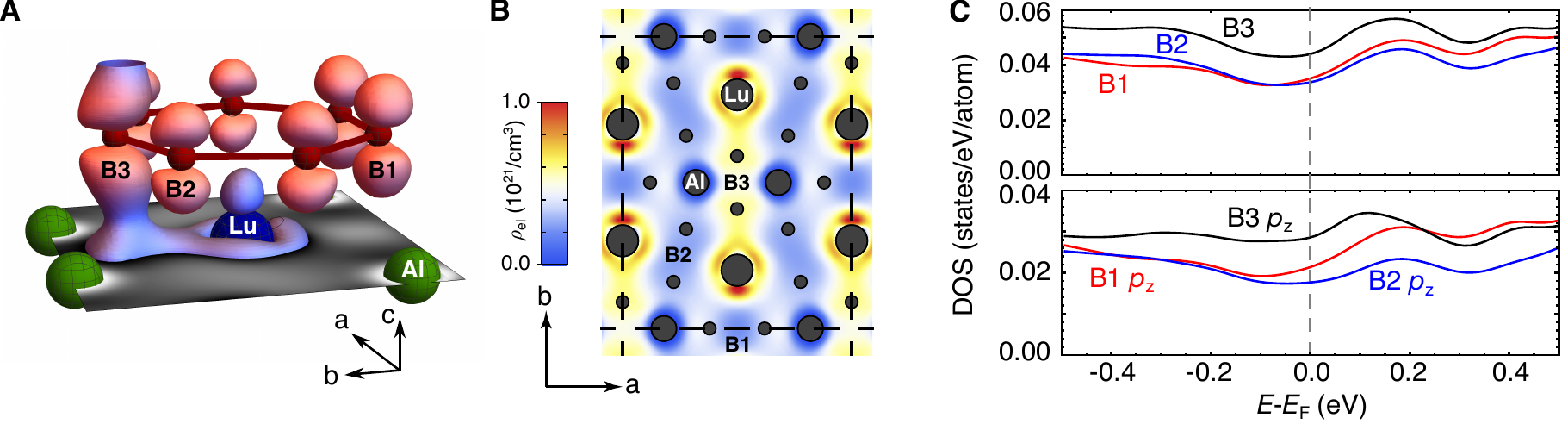}
	\end{overpic}
	\caption{Inequivalence of the boron atoms. (A) Charge carrier distribution $\rho_\text{el}$ in the proximity of the Lu atom and the sevenfold B ring. Electronic states within $\Delta E = k_\text{B} \cdot \SI{300}{\kelvin}$ of the Fermi level were included. The closed surfaces encapsulate regions with $\rho_\text{el} > 0.55 \cdot 10^{21}e/\text{cm}^3$. (B) The charge carrier distribution in the Lu/Al layer of the unit cell, with the B atoms projected. (C) Comparison of the density of states contributions of the B atoms. Upper panel: total contribution, lower panel: $p_z$ states only. }
	\label{fig:BoronAnisotropy}
\end{figure*}

The second implication is most evident from the hole-like Fermi surface sheet of \bLAB, Fig.~\ref{fig:DFTResult}B: There is a strong anisotropy of the electronic structure within the ($ab$) plane. To trace its origin, we reiterate that the DOS at the Fermi level $E_\text{F}$ has predominant Lu and B character, as shown in Fig.~\ref{fig:DFTResult}A. It is therefore suggestive to attribute the anisotropy to the Lu-B hybridization. This is confirmed when returning to real space: in Fig.~\ref{fig:BoronAnisotropy}A and B, we show the computed charge distribution in the unit cell for states close to $E_\text{F}$. We find a much larger orbital overlap between the Lu $d_{z^2}$ and the B3 $p_z$ states, when compared to B1 and B2. A similar result is obtained when comparing the contributions of the B atoms to the DOS, Fig.~\ref{fig:BoronAnisotropy}C. Over a wide energy range around $E_\text{F}$, the B3 states contribute roughly $30\%$ more to the DOS than B1 and B2, which originates fully from their $p_z$ orbitals.

This inequivalence of the B atoms cannot be assigned to the aforementioned atomic orbitals, since B $p_z$ and Lu $d_{z^2}$ orbitals are rotationally invariant along the crystal $c$ axis. Hence, no anisotropy is expected for an isolated stack of Lu atoms and ideal B rings. Consequentely, we must link it to the variation in Lu-B bond lengths (where the Lu-B3 bond length is the shortest), and/or the global anisotropy of the charge distribution in the orthorhombic unit cell. From our experiment, we cannot distinguish between these scenarios. However, as shown in Fig.~\ref{fig:BoronAnisotropy}B, we can infer that the charge carrier concentration is the largest within the Lu-B3-B3-Lu structure along the crystal $b$ axis. In contrast, the B1 and B2 atoms can be projected onto low-density stripes connecting the Al atoms along the same axis. The Al sites are effectively disconnected from charge carriers occupying states close to $E_\text{F}$, which marks them as ideally suited for chemical doping minimizing quasiparticle scattering.

We conclude by discussing the implication of this electronic anisotropy on \bYAB. As it is isostructural to \bLAB, and since the variations in Lu/Yb-B bond lengths are nearly identical (4\% in critical \bYAB, 5\% in non-critical \aYAB, and 6\% in non-critical \bLAB\ \cite{Macaluso2007}), we would expect that a similar variation of orbital overlap, and thus an anisotropic Kondo hybridization, should prevail in \bYAB. Yet, previous studies found an in-plane isotropic Kondo hybridization in $\alpha$- and \bYAB\ \cite{Kuga2019,Takano2016}. Our finding therefore provides evidence that the \emph{isotropic} Kondo hybridization and intrinsic quantum criticality in \bYAB\ emerge from an \emph{anisotropic} electronic structure \cite{Okawa2010,Watanabe2010,Kuga2018,Watanabe2019}. Identifying a mechanism which can counteract this electronic anisotropy and/or (self-)sta\-bi\-li\-zes the quantum critical phase remains an open problem.

\paragraph*{Acknowledgements.}
We thank Swee K. Goh for experimental assistance, and Hisatomo Harima for insightful discussions. PR acknowledges support from the Cusanuswerk (Germany) and the Oxford Quantum Materials Platform Grant (EPSRC No. EP/M020517/1). SF, FMG and MS acknowledge support from EPSRC Grants No. EP/N026691/1 and EP/K012894/1. This work is partially supported by Grants-in-Aids for Scientific Research on Innovative Areas (15H05882 and 15H05883), by Grants-in-Aid for Scientific Research (19H00650) from JSPS, and by CREST (JPMJCR18T3), Japan Science and Technology Agency (JST).

\bibliography{bib}

\end{document}